\documentclass[
aip,
amsmath,
amssymb,
reprint, 
]{revtex4-1}

\usepackage{graphicx}
\usepackage{booktabs}
\usepackage{tabularx}
\usepackage{hyperref}
\hypersetup{
	colorlinks=true,
	linkcolor=blue,
	filecolor=magenta,      
	urlcolor=blue,
	citecolor=black,
}

\usepackage{fancyhdr}
\fancypagestyle{diststmt}{%
	\fancyhf{}%
	\fancyfoot[C]{\footnotesize Distribution Statement A. Approved for public release: Distribution is unlimited.}%
}

\draft 
\begin{document}



\title{Sheath thickness measurements with the biased plasma impedance probe: Agreement with Child–Langmuir scaling %
}



\author{J. W. Brooks}
\email{john.w.brooks197.civ@us.navy.mil}
\altaffiliation{U.S. Naval Research Laboratory, Washington DC}

\author{R. Dutta}
\email{rdutta46@gatech.edu}
\altaffiliation{Georgia Institute of Technology, Atlanta GA}



\date{\today}

\begin{abstract}

Plasma sheaths play a central role in plasma–surface interactions, yet their thickness remains challenging to measure experimentally. Although classical analytical models such as the Child–Langmuir (CL) sheath model provide clear predictions for sheath thickness, experimental validation has been limited because most diagnostics either rely on indirect, multi-step inference (e.g., Langmuir probes) or require invasive and technically demanding techniques.
In this work, we demonstrate that the plasma impedance probe (PIP), when operated with a controlled DC bias, enables relatively direct, model-informed measurements of sheath thickness that are reasonably straightforward to implement experimentally. Across a range of discharge conditions, biased-PIP sheath thickness measurements are found to follow CL scaling closely, requiring a single, consistent empirical correction factor of $\alpha \approx 0.74$ to reconcile the measured thickness with CL predictions. Concurrent measurements of plasma density and electron damping show that probe biasing does not significantly perturb the bulk plasma density, supporting the validity of the biased-PIP approach.
Building on this validation, we leverage the empirically determined $\alpha$ factor to extend the floating (unbiased) PIP analysis to obtain model-dependent estimates of electron temperature and plasma potential without electrical biasing. A side-by-side comparison demonstrates close agreement between floating-PIP results and those obtained from a biased Langmuir probe. Taken together, these results establish the PIP as a complementary diagnostic to the Langmuir probe, expanding the range of accessible plasma measurements while providing experimental support for classical sheath models.
\end{abstract}


\maketitle 
\thispagestyle{diststmt}

\section{Introduction \label{sec:intro}}

Plasma sheaths are non-neutral boundary layers that form wherever a plasma contacts a material surface and play a central role in governing the exchange of charge, energy, and particles between plasmas and their boundaries \cite{Franklin2003,Hershkowitz2005,Hutchinson2002,Lieberman2005}. As a result, sheath physics is foundational to a wide range of laboratory and applied plasma systems, including semiconductor plasma processing \cite{Oehrlein2024}, electric propulsion devices \cite{Goebel2008}, and spacecraft–plasma interactions \cite{Garrett1981}. Although the basic structure of plasma sheaths has been studied extensively, authoritative reviews emphasize that experimentally validating key sheath properties under realistic conditions remains an ongoing challenge \cite{Oksuz2002,Hershkowitz2005}. In particular, accurate experimental measurements of sheath thickness continue to be difficult to obtain, limiting direct tests of classical sheath models.

The Child--Langmuir (CL) sheath model provides the canonical analytical benchmark for collisionless sheath thickness. Originally developed for space-charge-limited vacuum diodes, the model treats the sheath as a one-dimensional, steady-state region depleted of electrons and predicts a power-law dependence of sheath thickness on the sheath potential drop \cite{Franklin2003,Hutchinson2002,Lieberman2005}. In physical units, this dependence is commonly expressed as $t \propto V^{3/4}$. The Bohm criterion further constrains sheath formation by requiring ions to enter the sheath at or above the ion sound speed in order to sustain a monotonic potential profile \cite{Riemann1991}, and Hutchinson extended the CL framework by incorporating the Bohm current at the sheath edge \cite{Hutchinson2002,Goebel2008}. While these refinements modify the sheath entrance physics, the resulting sheath thickness scaling asymptotically approaches the classical CL form in the limit of large applied bias.

Experimental attempts to measure sheath thickness have employed a wide range of diagnostic techniques \cite{Franklin2003,Hershkowitz2005}, each involving trade-offs between experimental complexity, model dependence, and measurement directness. The most commonly used approach is the Langmuir probe (LP) \cite{Hutchinson2002,Faudot2019}, which does not measure sheath thickness directly but instead infers it by combining independently extracted plasma parameters with an assumed sheath model. As a result, LP-based sheath thickness estimates are inherently indirect and can deviate significantly from actual sheath dimensions when non-ideal effects are present, including non-Maxwellian electron distributions and sheath expansion \cite{Faudot2019}.  
Other techniques attempt to access sheath structure more directly but introduce their own limitations. Optical emission and ``dark space'' methods estimate sheath thickness from luminous plasma boundaries but are qualitative and limited by line-of-sight averaging and ambiguity in defining the sheath edge \cite{Mutsukura1990,Qin2006}. More direct approaches, such as emissive probe scans, laser-induced fluorescence, and electric-field mapping techniques, can provide detailed sheath information but typically require sophisticated instrumentation and are impractical for routine or embedded measurements \cite{Hershkowitz2005,Oksuz2002,Vorobiov2025,Ekanayaka2025}.  
More recently, RF impedance-based diagnostics, including the floating cutoff probe (CP), have emerged as a promising intermediate approach \cite{Kim2014,Han2014}. Interpreted using simple vacuum-sheath circuit models, these techniques yield relatively direct estimates of sheath thickness with modest hardware requirements and have been shown to exhibit general consistency with Child--Langmuir scaling, albeit with systematic offsets \cite{Kim2014}. While the CP shares conceptual similarities with the plasma impedance probe (PIP), the PIP’s ability to operate under controlled electrical bias, as discussed in the following section, enables more rigorous tests of sheath models and sheath thickness predictions.

The floating plasma impedance probe (PIP) is a diagnostic technique that characterizes plasma properties using impedance spectroscopy to measure the electrostatic RF response of a small electrode immersed in the plasma \cite{Balmain1964,Balmain1966,Grard1965,Blackwell2005RSI,Brooks2023,Brooks2024}. Its broadband impedance spectrum exhibits two characteristic resonances: a lower-frequency sheath resonance associated with the plasma--sheath boundary and a higher-frequency damped-plasma resonance related to the bulk plasma density.
The PIP was originally developed for ionospheric applications and deployed on rockets and satellites to measure electron density in situ \cite{Grard1965}, where relatively low plasma densities resulted in modest operating frequencies and calibration requirements.  
Adapting the PIP to higher-density laboratory plasmas, where plasma frequencies and associated RF challenges are substantially greater, required significant methodological advances. A key milestone was the work of Blackwell, who demonstrated floating-PIP operation under controlled laboratory conditions, introduced modern RF calibration techniques, and extracted plasma density from impedance spectra using an equivalent circuit model \cite{Blackwell2005RSI}. More recently, Brooks developed a low-dimensional theoretical framework for the PIP, closely related to Blackwell's work, and derived analytical expressions describing the sheath and damped-plasma resonances \cite{Brooks2023,Brooks2024}. Unlike earlier implementations that relied on identifying individual resonance features, this approach employs a single broadband, fittable expression that simultaneously yields plasma density, electron damping, and sheath thickness. By leveraging information across a wide frequency band and reducing sensitivity to noise and analysis artifacts, this framework substantially improves measurement precision when combined with appropriate calibration and experimental practices \cite{Brooks2024}. Taken together, these developments highlight the PIP as a mature and evolving laboratory diagnostic with expanding capability.

An important extension of the floating plasma impedance probe (PIP) is the introduction of electrical biasing, in which the probe electrode is driven with a controllable DC bias in addition to a small, superimposed RF excitation. Electrical biasing has long been employed in plasma diagnostics, most notably in Langmuir probes, and has also been applied to several RF-based techniques to modify sheath conditions and extend their operating range \cite{Pandey2020}. A key early demonstration of the biased PIP was provided by Blackwell, who observed that the sheath resonance shifted systematically to lower frequencies with increasingly negative probe bias, consistent with expectations from Child--Langmuir scaling \cite{Blackwell2005PoP}. Although that study did not directly extract sheath thickness from the resonance behavior, it established a quantitative link between probe bias, sheath structure, and the measured impedance response.  
Building on this foundation, Walker and co-workers developed analysis techniques that extended the capabilities of the biased PIP by demonstrating how plasma potential and the electron energy distribution function could be inferred from biased RF impedance spectra \cite{Walker2008,Walker2010}. More recently, Boris validated these methods experimentally in RF plasmas, confirming the sensitivity of the biased PIP to sheath-related phenomena \cite{Boris2012}. Collectively, these studies established the biased PIP as a valuable tool for probing sheath and plasma properties, while also highlighting that sheath thickness estimates were obtained indirectly through resonance behavior, thereby motivating the more direct sheath thickness measurement pursued in the present work.

Despite the maturity of analytical sheath models, a clear diagnostic gap persists: sheath theory requires experimental validation, yet existing measurement techniques are often indirect, highly model dependent, or impractical for routine use. In this work, we address this gap by combining the biased plasma impedance probe (PIP) with a recently developed PIP model to obtain a relatively direct measurement of sheath thickness. Section~\ref{sec:theory} reviews Child--Langmuir (CL) sheath theory and the theoretical framework underlying the PIP. Section~\ref{sec:exp_setup} describes the experimental setup and the implementation of a hybrid diagnostic that integrates Langmuir probe and PIP operation within a single probe assembly. In Section~\ref{sec:biased_pip}, we demonstrate that the biased PIP yields sheath thickness measurements in close agreement with CL theory across a range of probe biases. Section~\ref{sec:floating_pip} extends the floating PIP (i.e., without applied bias) to enable model-based measurements of electron temperature and plasma potential in addition to established measurements of density, sheath thickness, and electron damping. Finally, Section~\ref{sec:conclusions} summarizes the conclusions.

\section{Theory  \label{sec:theory}}

The Child--Langmuir (CL) sheath model provides the canonical analytical scaling for sheath thickness by treating the sheath as a one-dimensional, planar, collisionless, space-charge-limited region with cold ions injected at rest (i.e., the Bohm criterion is not enforced)~\cite{Hutchinson2002,Lieberman2005}. Solving Poisson’s equation under these assumptions yields the CL sheath thickness,
\begin{equation}
	t_{\mathrm{CL}} = \frac{2}{3}\,\lambda_D \left( \frac{-2\Delta V}{T_{eV}} \right)^{3/4},
	\label{eq:t_cl}
\end{equation}
where the Debye length is given by
\begin{equation}
	\lambda_D = \left( \frac{\varepsilon_0 T_{eV}}{n_e e} \right)^{1/2},
	\label{eq:lambda_d}
\end{equation}
and
\begin{equation}
	\Delta V = V_{\mathrm{probe}} - V_{\mathrm{plasma}}
	\label{eq:delta_v}
\end{equation}
is the sheath potential defined relative to the probe voltage. Note that the sign of $\Delta V$ is opposite of convention. In these expressions, $n_e$ is the plasma density and $T_{eV} = k_B T_e / e$ is the electron temperature expressed in electron volts.
Despite its simplifying assumptions, the CL law remains the baseline analytical sheath model, as more sophisticated treatments (e.g.,  Hutchtinson's model) often reduce to this same scaling in the large $|\Delta V|/T_{eV}$ limit \cite{Hutchinson2002}. In practice, Langmuir probe measurements are commonly used to provide the values of $n_e$, $V_{\mathrm{plasma}}$, and $T_{eV}$ required to evaluate Eq.~\ref{eq:t_cl}.

We now summarize the plasma impedance probe (PIP) model developed by Brooks, which is presented in detail elsewhere \cite{Brooks2023,Brooks2024} and is based on earlier work by Blackwell \cite{Blackwell2005RSI,Blackwell2020}. The model assumes an electrostatic response and spherical geometry, treating the vacuum--sheath region and the plasma region as two spherical, concentric capacitors in series. The vacuum--sheath capacitor has a finite thickness $t_{\mathrm{pip}}$ and a vacuum dielectric constant $\varepsilon_0$.
The plasma region is assumed to be capacitor filled with a homogeneous, cold, damped, and unmagnetized plasma dielectric, $\varepsilon_0 \varepsilon_p(\omega)$, where
\[
\varepsilon_p(\omega) = 1 - \frac{\omega_p^2}{\omega(\omega - j\nu)},
\]
$\nu$ is a generalized electron damping (or loss) term that includes both collisional and collisionless contributions, and 
\[
\omega_p = \sqrt{\frac{n_e e^2}{\epsilon_0 m_e}} 
\]
is the electron plasma frequency.  

The resulting model for the PIP electrical impedance may be written in normalized form as
\begin{align}
	Z'_{\mathrm{pip}}(\omega)
	&\equiv \frac{Z_{\mathrm{pip}}(\omega)}{Z_m} \notag \\
	&= \frac{\nu' \omega' (1 - t'_{\mathrm{pip}})
		- j(\omega'^2 - \omega_-^2)(\omega'^2 - \omega_+^2)}
	{\omega'\!\left[(\nu' \omega')^2 + (\omega'^2 - 1)^2\right]}
	\label{eq:Zpip}
\end{align}
where $\omega' = \omega / \omega_p$ is the normalized angular frequency, $\nu' = \nu / \omega_p$ is the normalized electron damping rate,  $t_{\mathrm{pip}}$ is the sheath thickness,
\[
t'_{\mathrm{pip}} = \frac{t_{\mathrm{pip}}}{t_{\mathrm{pip}} + r_{\mathrm{pip}}} \in [0,1)
\]
is normalized sheath thickness, $r_{\mathrm{pip}}$ is the PIP-monopole's radius, and
\[
Z_m = \left(4\pi \varepsilon_0 r_{\mathrm{pip}} \omega_p \right)^{-1}
\]
is the characteristic impedance associated with the spherical monopole geometry at the plasma frequency.    

Equation~\ref{eq:Zpip} is characterized by two resonances ($\omega_\pm'$), defined by the roots of $\mathrm{Im}\!\left\{ Z'_{\mathrm{pip}}(\omega) \right\} = 0.$
These resonance frequencies take the form,

\begin{equation}
	\begin{aligned}
	\omega_\pm' &\equiv \frac{\omega_\pm}{\omega_p} \\
	&=\sqrt{\frac{1+t'_{\mathrm{pip}}-\nu'^2
			\pm \sqrt{(1+t'_{\mathrm{pip}}-\nu'^2)^2-4t'_{\mathrm{pip}}}}{2}}.
	\end{aligned}
	\label{eq:omega_pm}
\end{equation}
In the low-damping limit, these expressions reduce to
\begin{subequations}
	\label{eq:omega_limits}
	\begin{align}
		\lim_{\nu'^2 \rightarrow 0} \omega_- &= \sqrt{t'_{\mathrm{pip}}}\,\omega_p,
		\label{eq:omega_minus_lim} \\
		\lim_{\nu'^2 \rightarrow 0} \omega_+ &= \omega_p.
		\label{eq:omega_plus_lim}
	\end{align}
\end{subequations}
We refer to the lower resonance $\omega_-$ as the \emph{sheath resonance}, since it depends explicitly on the normalized sheath thickness $t'_{\mathrm{pip}}$, consistent with previous experimental observations \cite{Kim2014,Blackwell2005RSI}. 
Combining the sheath thickness dependence of $\omega_-$ (Eq.~\ref{eq:omega_minus_lim}) into Eq.~\ref{eq:t_cl},
we therefore expect $\omega_- \propto -|\Delta V|^{3/8}$.
We refer to the upper resonance $\omega_+$ as the \emph{damped-plasma resonance}, as finite electron damping shifts this resonance below the plasma frequency. Because $\omega_+$ is independent of sheath thickness in the low-damping limit (Eq.~\ref{eq:omega_plus_lim}), it is expected to remain insensitive to changes in sheath thickness induced by variations in probe bias $\Delta V$.
We justify our use of the low-damping approximation in this work as we measured $0.01 < \nu'^2 < 0.05$ for $\Delta V < 0$ in all the plasma conditions studied here.

As a quick aside, we use the reflection coefficient, 
\begin{equation}
	\Gamma(\omega) = \frac{Z(\omega) - 50 \Omega}{Z(\omega) + 50 \Omega}
	\label{eq:Gamma}
\end{equation}
for numerical fitting and data analysis, where 50 $\Omega$ is the characteristic impedance of the measurement system. We adopt this representation because prior work has shown that analysis in $\Gamma$ provides a substantial improvement in signal-to-noise ratio~\cite{Brooks2024}.  Combining Eqs.~\ref{eq:Zpip} and~\ref{eq:Gamma} results in a $\Gamma_{pip}(\omega)$ model that retains the same resonance frequencies as the impedance model (Eqs.~\ref{eq:omega_pm} and \ref{eq:omega_limits}).

Finally, fitting calibrated PIP measurements to $Z_{pip}(\omega)$ or $\Gamma_{pip}(\omega)$ provides reasonably direct measurements of four plasma properties:  $\omega_p$, $n_e$, $t_{\mathrm{pip}}$ and $\nu$.  And because $t_{\mathrm{pip}}$ is linked to the electron temperature $T_e$ through the sheath thickness (e.g., Eq.~\ref{eq:t_cl}), this quantity can be extracted with the inclusion of additional EEDF and sheath assumptions~\cite{Georgin2024}. 

Finally, we highlight an important distinction between the sheath thickness predicted by the Child--Langmuir (CL) model and the sheath thickness inferred from the plasma impedance probe (PIP). Although both models employ an idealized step-function electron density profile, they arrive at this abstraction for fundamentally different reasons. In the CL model, the sheath is treated as an electron-depleted, space-charge-limited region whose thickness is determined by planar, one-dimensional electrostatics and ion space-charge physics. In contrast, the PIP model assumes a step-function electron density profile as part of a geometric, circuit-based abstraction in which a finite-thickness, spherical vacuum sheath is introduced to model the RF impedance of the plasma--probe boundary.
As a consequence of these related but differing assumptions, the sheath thicknesses $t_{\mathrm{CL}}$ and $t_{\mathrm{pip}}$ should not be expected to coincide exactly. 

To facilitate $t_{\mathrm{pip}} \neq t_{\mathrm{CL}}$, we introduce $\alpha$, a dimensionless scaling factor, to relate the two sheath thicknesses, 
\begin{equation}
	t_{\mathrm{pip}} \approx \alpha\, t_{\mathrm{CL}}
	= \alpha\,\frac{2}{3}\lambda_D
	\left(\frac{-2\Delta V}{T_{eV}}\right)^{3/4},
	\label{eq:alpha_scaling}
\end{equation}
This relation is constructed with a zero intercept, ensuring that both sheath thicknesses vanish as $V_{\mathrm{probe}} \rightarrow V_{\mathrm{plasma}}$.

\begin{figure}
	\includegraphics{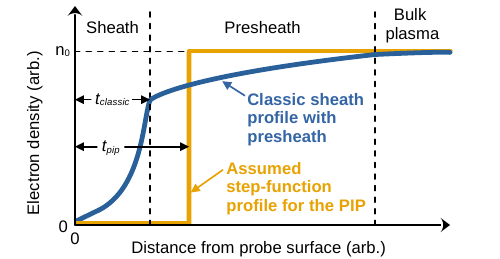}
	\caption{
		Conceptual comparison of electron density profiles near a probe surface.
		The PIP model assumes a step-function electron density profile corresponding to a vacuum sheath of thickness $t_{\mathrm{pip}}$.
		In contrast, more complete classical descriptions include a continuous sheath profile and an extended quasi-neutral presheath region \cite{Lieberman2005}.
		The figure is intended to provide physical intuition regarding the differing levels of model abstraction rather than a quantitative comparison.
	}
	\label{fig:sheath_profiles}
\end{figure}

We note in passing that more complete classical sheath descriptions include an extended quasi-neutral presheath region and a continuous electron density profile governed by Boltzmann statistics \cite{Lieberman2005}. Such models describe a broader plasma boundary layer than either the idealized CL or PIP sheath abstractions. Figure~\ref{fig:sheath_profiles} illustrates these conceptual differences by comparing the step-function electron density profile assumed by the PIP model with a representative classical sheath profile that includes a presheath region. While direct quantitative comparison between these more complete sheath models and the PIP formulation is beyond the scope of the present work, this comparison provides useful physical context for later discussions of presheath effects and motivates future extensions of impedance-based sheath modeling.

\section{Experimental setup  \label{sec:exp_setup}}

The plasma source used in this experiment was a custom 20~A class thermionic hollow cathode (HC) \cite{Georgin2021} operated with an argon flow rate of 30~sccm. The HC was mounted inside a cylindrical vacuum chamber 0.5~m in diameter and 1~m in length, which was evacuated by a CTI400 cryopump. Under typical operating conditions, the chamber pressure was maintained in the range of 100--200~\textmu Torr, which facilitated a relatively low plasma damping rate. Experiments were conducted at HC discharge currents of 6, 8, 10, 15, and 20~A in order to span a range of plasma densities.

\begin{figure}
	\includegraphics{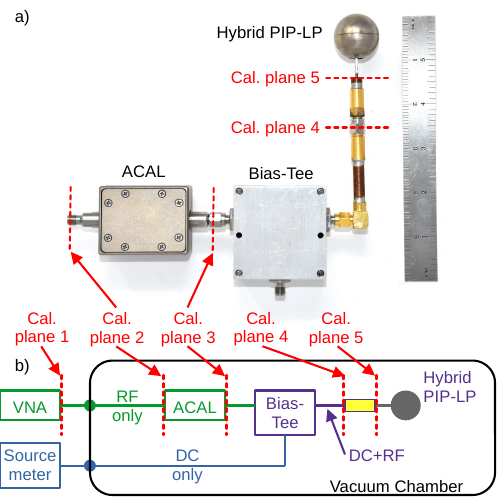}
	\caption{(a) Photograph of the 1-inch spherical probe and adjacent electrical hardware used in the hybrid PIP--LP diagnostic. (b) Schematic of the electrical system, consisting of three circuit regions: RF-only (green), DC-only (blue), and a superimposed DC+RF (purple). These circuits enable operation of the same probe in three modes: LP, floating PIP, and biased PIP. Calibration planes used in the RF de-embedding procedure are indicated.  }
\label{fig:pip_setup}
\end{figure}

Figure~\ref{fig:pip_setup} illustrates the probe and supporting electrical hardware used in this experiment. The hybrid probe supports three distinct modes of operation: a spherical Langmuir probe (LP), a floating PIP (plasma impedance probe), and a biased PIP. The probe consists of a 1-inch-diameter stainless steel sphere attached to the inner conductor of a stripped RG401 semi-rigid coaxial cable using set screws. Also shown are the adjacent RF components and transmission lines that form part of the measurement and calibration chain. Key to this hybrid architecture is that this single probe configuration can be operated in each diagnostic mode without breaking vacuum.  The probe was placed roughly one foot axially downstream of HC's gridded, annular anode.

Figure~\ref{fig:pip_setup}b shows the electrical circuit supporting hybrid PIP--LP operation. The system consists of three distinct circuit regions: an RF-only region used to operate the PIP, a DC-only region used to bias the probe, and a combined DC+RF region that superimposes the DC and RF signals, which enables all three modes of operation. A bias tee (Mini-Circuits Z3BT-2R15G+) is used to superimpose the DC and RF signals while maintaining isolation between the two circuits. This overall configuration closely follows that employed in prior biased-PIP experiments \cite{Blackwell2005PoP}.

The DC electrical circuit notably consisted of a Keysight 2560 sourcemeter, which operated the probe both as a traditional swept LP and as the bias supply for biased-PIP operation. During LP measurements, the sourcemeter swept the probe voltage from $-20$~V to $+99$~V using 200 uniformly spaced points, with the integration time set to one power-line cycle (NPLC~=~1) to minimize 60~Hz line noise. The resulting current--voltage characteristics were analyzed using the spherical-probe method described by Lobbia \cite{Lobbia2017}, following steps~1--7 to extract the electron temperature, plasma density, and plasma potential from the electron current. These quantities were then used in Eq.~\ref{eq:t_cl} to calculate the Child--Langmuir sheath thickness.
For completeness, the ion current was also analyzed using the thin-sheath, thick-sheath (OML), and transitional sheath models described by Lobbia \cite{Lobbia2017}. However, the electron-current-based analysis was adopted for consistency across all discharge conditions.

In addition to the probe and bias tee, the RF electrical circuit consisted of a Keysight E5071C vector network analyzer (VNA), a Copper Mountain ACMP-2506-112 programmable auto-calibration (ACAL) unit, and five calibration planes. The VNA measured the raw reflection coefficient $\Gamma$ of the RF circuit and probe using linear frequency sweeps from 1~MHz to 1~GHz with 1601 frequency points. These complex measurements were subsequently calibrated (as described below) and fit using the PIP model, $\Gamma_{\mathrm{pip}}(\omega)$, to extract the plasma density, sheath thickness, and electron damping \cite{Brooks2023,Brooks2024}.
During biased-PIP operation, the sourcemeter supplied a DC bias to the probe over the range $-20$~V to $+90$~V in 5~V increments while RF measurements were performed. For floating-PIP operation, the sourcemeter output was disabled, allowing the probe to float electrically while RF measurements were performed.

The RF electrical system between the VNA and the probe introduces additional impedance that must be removed through calibration (de-embedding) in order to isolate the probe’s intrinsic response. Calibration is arguably the most technically demanding aspect of operating a PIP. While only a brief summary is provided here, detailed descriptions of the procedure may be found in Refs.~\cite{Brooks2023,Brooks2024,Tejero2025}. This calibration employs five calibration planes (Fig.~\ref{fig:pip_setup}) and three successive calibration steps, implemented using de-embedding routines from the \texttt{scikit-rf} (SKRF) library \cite{Arsenovic2022}.
In Step~1 (planes~1 to 2), the VNA measured the three RF standards contained in the in-situ ACAL unit immediately prior to each PIP measurement. These measurements were combined with previously acquired ``truth'' measurements of the same standards in a one-port calibration routine. In Step~2 (planes~2 to 4), a two-port de-embedding routine was applied using previously acquired two-port measurements between these planes. In Step~3 (planes~4 to 5), the probe stem was calibrated by subtracting a two-port, lossless transmission-line model of length $L$. The value of $L$ was determined by combining the transmission-line model with the vacuum PIP model, measuring the PIP, and fitting the combined model to the plane-4-calibrated vacuum measurement.
Together, this three-step calibration procedure removes the contributions of external hardware (planes 1 to 5) and isolates the broadband response of the probe for direct comparison with the PIP model.

At each of the five hollow-cathode discharge currents, three types of measurements were performed: a swept LP measurement, a floating PIP measurement, and a biased-PIP sweep. Because the biased-PIP sweep required several minutes to complete, the LP and floating-PIP measurements were acquired both immediately before and immediately after the biased-PIP sweep. This bracketing procedure was used to verify that the plasma conditions remained stable throughout the measurement sequence.

\section{Biased PIP Results and Discussion  \label{sec:biased_pip}}

\subsection{Bias-dependent PIP spectra and resonance behavior \label{subsec:biased_pip_resonances}}

\begin{figure}[t]
	\centering
	\includegraphics{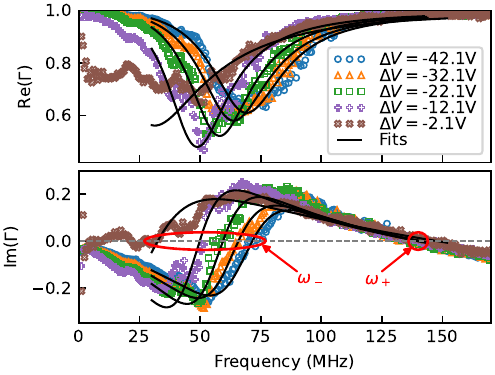}
	\caption{Calibrated PIP measurements for the $I_D = 10$~A discharge case at select probe biases, showing the real (top) and imaginary (bottom) components of the reflection coefficient, $\Gamma$, along with corresponding fits from the PIP model.  The lower, sheath resonance ( $\omega_-$) is shown to be a function of $\Delta V$.  In contrast, the upper, damped-plasma resonance ( $\omega_+$) is not. }
	\label{fig:biased_pip_spectra} 
\end{figure}

Figure~\ref{fig:biased_pip_spectra} presents example calibrated measurements of the plasma impedance probe (PIP) at several probe biases for the $I_D = 10$~A discharge case. The plasma potential, $V_{\mathrm{plasma}}$, was obtained from the Langmuir probe (LP) analysis, while the probe bias, $V_{\mathrm{probe}}$, was set by the sourcemeter, such that $\Delta V = V_{\mathrm{probe}} - V_{\mathrm{plasma}}$. The plots show both the real and imaginary components of the fully calibrated measurements, $\Gamma$, along with fits from the PIP model, where the PIP model is simultaneously fit to both the real and imaginary components of the measurement as described in prior work~\cite{Brooks2023,Brooks2024}. The close agreement between the measured spectra and the model fits demonstrates, qualitatively, that the PIP model accurately captures the probe’s broadband impedance response. 
All plasma properties reported in the following sections, including density, sheath thickness, and electron damping, are similarly obtained directly from these broadband fits of the PIP model.

Several notable results emerge from the data shown in Fig.~\ref{fig:biased_pip_spectra}. First, the damped-plasma resonance, $\omega_+$, remains approximately constant across all probe bias voltages. This behavior is consistent with Eq.~\ref{eq:omega_plus_lim}, which predicts that $\omega_+$ is independent of sheath thickness, and suggests that biasing the probe does not significantly perturb the bulk plasma density. 

Second, the sheath resonance, $\omega_-$, shifts systematically to lower frequencies with increasingly negative probe bias. This trend is in agreement with Eq.~\ref{eq:omega_minus_lim} and with prior observations reported by Blackwell \cite{Blackwell2005PoP}, where larger negative bias corresponds to a thicker sheath and, consequently, a reduced sheath-resonance frequency.

Third, for small negative biases ($0 \gtrsim \Delta V \gtrsim -10$~V), the spectra develop additional low-frequency features ($\omega/2\pi \lesssim 30$~MHz) that are not captured by the PIP model. Measurements in this regime are therefore less reliable and fall outside the model’s assumptions, potentially due to plasma nonuniformity, sheath-edge gradients, or partial sheath collapse near the floating condition. Accordingly, all fits reported in this work are restricted to frequencies $\omega/2\pi > 30$~MHz.

Finally, we do not observe the additional $\Delta V < 0$ resonance reported by Blackwell \cite{Blackwell2005PoP}, which was attributed to the presheath region. The absence of this feature in the present measurements may reflect differences in plasma conditions, probe geometry, or diagnostic sensitivity.

For completeness, measurements were also examined in the ion-sheath regime ($\Delta V > 0$). As expected, the vacuum-sheath assumption underlying the PIP model is not valid in this regime, and the corresponding fits do not adequately capture the measured spectra (i.e., the fits were terrible as physically important features were not present in the measurements). Accordingly, the present analysis is restricted to the electron-sheath regime ($\Delta V < 0$). Although the damped-plasma resonance, $\omega_+$, was observed for $\Delta V > 0$, we did not observe the additional ion-sheath-related resonance reported by Blackwell \cite{Blackwell2005PoP}. 

\subsection{Comparison of biased-PIP sheath thickness with Child–Langmuir theory  \label{subsec:biased_pip_sheath_thick}}

\begin{figure}[t]
	\centering
	\includegraphics{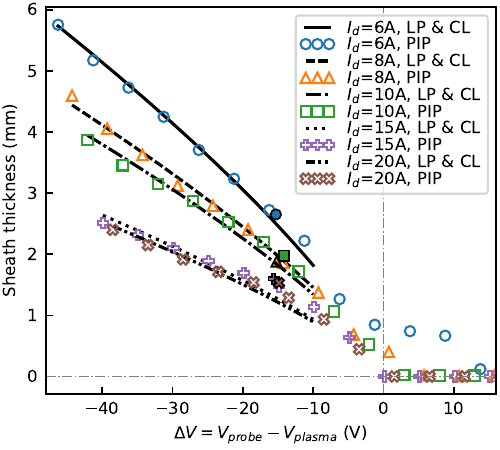}
	\caption{Comparison of biased-PIP sheath thickness measurements, $t_{\mathrm{pip}}$, with Child--Langmuir sheath thicknesses, $t_{\mathrm{CL}}$, calculated from Langmuir-probe measurements of density and temperature and scaled by $\alpha = [0.71, 0.74, 0.74, 0.74, 0.78]$ for the five discharge currents, respectively. Empty markers denote biased-PIP measurements, while filled markers denote floating-PIP measurements ($\Delta V \approx -15$~V). The results show consistent agreement between $t_{\mathrm{pip}}$ and $\alpha t_{\mathrm{CL}}$ across all discharge conditions.}
	\label{fig:tpip_vs_tcl}
\end{figure}

Figure~\ref{fig:tpip_vs_tcl} compares the biased-PIP and floating-PIP sheath thickness measurements, $t_{\mathrm{pip}}$, with Child--Langmuir sheath thicknesses, $t_{\mathrm{CL}}$. The $t_{\mathrm{CL}}$ values were calculated using Langmuir-probe measurements of plasma density and electron temperature and scaled by a constant factor $\alpha$ (see Eq.~\ref{eq:alpha_scaling}). Results are shown for all five discharge currents (6, 8, 10, 15, and 20~A). 

The value of $\alpha$ was determined for each discharge current by a least-squares fit of Eq.~\ref{eq:alpha_scaling} to the $t_{\mathrm{pip}}$ measurements over the range $\Delta V < -10$~V. The resulting values, $\alpha = [0.71, 0.74, 0.74, 0.74, 0.78]$, exhibit a mean of $\alpha \approx 0.74$ with a standard deviation of $\sigma_\alpha \approx 0.02$. 
The close agreement between $t_{\mathrm{pip}}$ and $\alpha t_{\mathrm{CL}}$ across all discharge currents, together with the consistency of $\alpha$, suggests that the linear relationship introduced in Eq.~\ref{eq:alpha_scaling} provides a robust empirical mapping between the two sheath models. A similar comparison using Hutchinson's sheath thickness model~\cite{Hutchinson2002} (not shown) yielded a consistent scaling factor of $\alpha \approx 0.46 \pm 0.01$ across all discharge currents.

Although Fig.~\ref{fig:tpip_vs_tcl} shows close agreement between $t_{\mathrm{pip}}$ and $\alpha t_{\mathrm{CL}}$, the physical origin of $\alpha \neq 1$ remains uncertain. 
One possibility is that $\alpha$ reflects systematic differences between the underlying sheath idealizations: the Child--Langmuir model assumes a \emph{planar} geometry, whereas the PIP model assumes a \emph{spherical} geometry. 
A second, closely related possibility is that $\alpha$ corrects for the difference between the CL's assumed \emph{thin} sheath and the PIP's assumed sheath of \emph{finite-thickness}.  For context, the PIP's sheath thickness measurements were consistently $t_{\mathrm{pip}}/r_{\mathrm{pip}} \lesssim 0.3$, and therefore curvature and finite-thickness effects of the sheath may have been appreciable.
A third possibility is that $\alpha$ compensates for systematic uncertainties in the Langmuir-probe-derived inputs used to compute $t_{\mathrm{CL}}$, since errors in $n_e$ and $T_{eV}$ propagate directly into the predicted sheath thickness. However, if measurement error were the dominant contributor, one would generally expect $\alpha$ to vary more strongly across discharge conditions, particularly given that PIP--LP density agreement is better at lower discharge currents and poorer at higher currents (see Sec.~\ref{subsec:biased_pip_n_and_T}). The observed consistency of $\alpha \approx 0.74$ across all cases therefore suggests that $\alpha$ is more likely capturing a systematic difference between the two sheath model forms rather than condition-dependent measurement error alone. Similar systematic offsets have been reported in related RF sheath diagnostics; for example, Kim \cite{Kim2014} found that cutoff-probe sheath thickness estimates tended to exceed Child--Langmuir predictions and attributed the discrepancy to oversimplifications of the vacuum-sheath model and to unmodeled effects such as radiative losses and wave--plasma interactions.

\subsection{Bias dependence of plasma density and electron damping \label{subsec:biased_pip_n_and_T}}

\begin{figure}[t]
	\centering
	\includegraphics{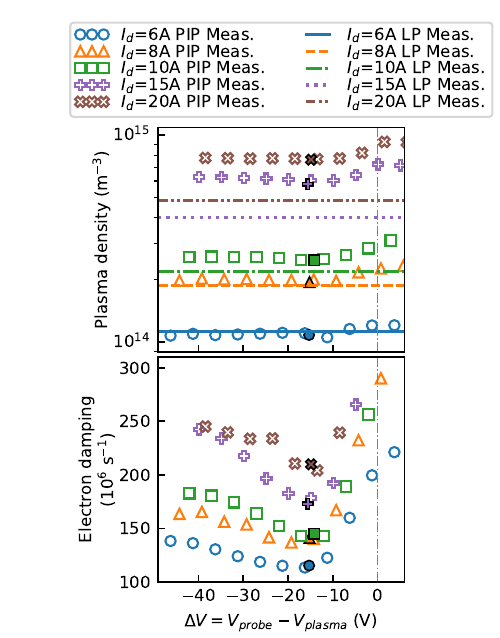}
	\caption{
		(a) Comparison of plasma density measurements obtained from the plasma impedance probe (PIP) and the Langmuir probe (LP) as a function of probe bias. Horizontal lines indicate the LP density measured at each discharge current, while symbols show PIP-derived densities. (b) Electron damping rate $\nu$ extracted from PIP fits as a function of probe bias. Empty markers denote biased-PIP measurements, and filled markers indicate floating-PIP measurements ($\Delta V \approx -15$~V).}
	\label{fig:density_damping}
\end{figure}

Fig.~\ref{fig:density_damping} presents the PIP's other two measurements: plasma density ($n_e$) measurements obtained of the floating-PIP and biased-PIP and compared with the LP, along with the PIP's measured electron damping rate ($\nu$). 
Together, these measurements provide additional context for interpreting the sheath-thickness results by illustrating how biasing influences other plasma parameters inferred from the PIP model.

Figure~\ref{fig:density_damping}a highlights two key trends. 
First, the PIP-derived electron density remains essentially constant with bias for $\Delta V \lesssim -10~\mathrm{V}$, indicating that neither DC biasing nor sheath-thickness variations significantly influence the bulk density measurement. 
This behavior is expected because the PIP density is extracted from the damped-plasma resonance near $\omega_p$, which is first-order insensitive to sheath structure (Eq.~\ref{eq:omega_plus_lim}) and does not rely on a Maxwellian electron energy distribution function (EEDF). 
Consistent with this interpretation, the present measurements show close agreement between floating and biased PIP densities across all discharge currents, suggesting that DC biasing does not introduce a systematic error in the PIP density measurement.

Second, while PIP- and LP-derived densities agree closely at lower discharge currents (6--10~A), systematic discrepancies emerge at higher currents. 
A likely explanation is the increasing susceptibility of LP analysis to surface effects (e.g., outgassing) at elevated discharge currents (i.e., elevated probe-surface temperatures), rather than a failure of the PIP measurement. 
Blackwell~\cite{Blackwell2015} demonstrated that \emph{biased} impedance-probe density measurements remain accurate even when the probe surface is intentionally contaminated, whereas plasma potential measurements, and by extension LP-derived density, are strongly affected by surface condition. 
An additional explanation may include possible non-Maxwellian distributions (e.g., beams) at the higher discharge currents, which notably influence LP analysis.  
Alternatively, the PIP-monopole effectively measures both near and far-field components but weights the near-field more heavily due to the $1/r^2$ electric field falloff~\cite{Brooks2023} and other loss effects.  Therefore, gradients in the plasma may have an outsized impact on the PIP.

The damping results in Fig.~\ref{fig:density_damping}b show a clear increase in the extracted electron damping parameter $\nu$ with increasing negative bias, which at first appears inconsistent with the expectation that $\nu$ is a bulk plasma property independent of sheath thickness. 
A possible explanation is that the PIP model treats the sheath and plasma regions' profile as a step function with a lossless dielectric in the sheath, whereas in reality elevated RF dissipation occurs in the sheath--presheath region. 
In this region, strong and spatially inhomogeneous RF electric fields can drive additional dissipative processes, such as stochastic electron heating \cite{Lieberman2005}. 
When the sheath is thin, the probe’s RF perturbation penetrates more deeply into the bulk plasma and the measured response may be dominated by bulk damping. 
As the sheath thickens with increasing bias, a larger fraction of the RF potential drop possibly becomes confined to the sheath--presheath region, where dissipation is stronger. 
As a result, the effective damping parameter $\nu$ inferred from the PIP measurement is a function of $\Delta V$, even if the intrinsic bulk plasma damping remains unchanged.

Finally, we briefly consider whether RF-only measurements from the biased PIP (i.e., without DC current or voltage measurements) could be extended to extract the plasma potential $V_{plasma}$ and electron temperature $T_{eV}$ by adapting methods developed by Walker and Boris~\cite{Walker2008,Walker2010,Boris2012,Blackwell2015}. 
In principle, this approach analyzes $Z_{\mathrm{pip}}(\omega)$ at intermediate frequencies between the ion and electron plasma frequencies ($\omega_{pi} \ll \omega \ll \omega_{pe}$), where electrons remain mobile while ions are effectively stationary. 
Under these conditions, the electron response can be isolated without explicitly subtracting an ion current, in contrast to conventional Langmuir probe analysis. 
In practice, however, these methods are highly sensitive to the voltage resolution of the applied bias. 
In the present experiment, the $5~\mathrm{V}$ step size used for the biased-PIP sweep was insufficient to enable accurate application of these techniques, and we therefore do not present such results here. 
We note, however, that future experiments designed with finer bias resolution and careful consideration of RF filter cutoff frequencies in the electrical circuit (e.g., within the bias tee) may enable this analysis, further expanding the capability of the biased PIP as a stand-alone diagnostic.

\section{Adding temperature and floating potential capability to the floating PIP  \label{sec:floating_pip}}

In many plasma systems, DC biasing of a probe is undesirable because it can significantly perturb the plasma~\cite{Boris2012,Walker2010}. 
While the \emph{biased} plasma impedance probe (PIP) enables direct access to sheath physics, it shares this limitation with traditional Langmuir probes (LP). 
In contrast, electrically floating probes (i.e., the floating PIP) are less sensitive to particle beams, surface contamination, and secondary electron emission, and, additionally do not require RF compensation~\cite{Sudit1994,Faudot2019}.  However, EEDF properties, such as temperature and plasma potential, are difficult to measure with in-situ electrical probes without biasing.  

However, the results from Sec.~\ref{sec:biased_pip} provide a unique opportunity to enable the floating PIP to measure temperature and plasma potential without biasing.  In this section, we develop the necessary theory, which requires introducing two equations to solve for our two unknowns.

Our first equation is Eq.~\ref{eq:alpha_scaling}, which is our empirical relation that explicitly connects the PIP’s sheath thickness measurement to the Child--Langmuir sheath model, which depends on our two unknowns ($T_{eV}$ and $V_{plasma}$), as well as quantities that the PIP provides ($n_e$ and $t_{pip}$), and the floating potential ($V_{float}$), which we can measure with our source meter. 

For our second equation, we adopt elements of Langmuir probe theory, where the floating condition is defined by the balance of ion and electron currents,
\begin{equation}
	I_i(V_\mathrm{float}) + I_e(V_\mathrm{float}) = 0 .
	\label{eq:current_balance}
\end{equation}
For the ion current $I_i$, we assume a spherical probe operating in the transitional sheath regime ($3 \lesssim r_\mathrm{probe}/\lambda_D \lesssim 50$), which interpolates between the thin-sheath and orbital-motion-limited (OML) limits~\cite{Lobbia2017,Narasimhan2001}.  We chose this model because our sheath thickness measurements puts us solidly in the middle of the this $r_\mathrm{probe}/\lambda_D$ range.
The electron current, $I_e$, is modeled assuming a Maxwellian electron energy distribution, quasineutrality at the sheath edge, and negligible secondary electron emission.
Under these assumptions, expanding and simplifying Eq.~\ref{eq:current_balance} yields
\begin{subequations}\label{eq:floating_system}
	\begin{align}
		\exp\!\left(\frac{\Delta V}{T_{eV}}\right)
		&= a \sqrt{\frac{m_e}{m_i}}
		\left(-\frac{\Delta V}{T_{eV}}\right)^{b},
		\label{eq:floating_system_a}
	\end{align}
	where the coefficients
	\begin{align}
		a
		&= 1.58 + \left(-0.056 + 0.816\,\frac{r_\mathrm{probe}}{\lambda_D}\right)^{-0.744},
		\label{eq:floating_system_b}\\[4pt]
		b
		&= -0.933 + \left(0.0148 + 0.119\,\frac{r_\mathrm{probe}}{\lambda_D}\right)^{-0.125},
		\label{eq:floating_system_c}
	\end{align}
\end{subequations}
account for the transitional sheath geometry through their dependence on $r_\mathrm{probe}/\lambda_D$.  This equation depends on the same knowns and unknowns as Eq.~\ref{eq:alpha_scaling} but is arrived at from a fundamentally differently approach and with a \emph{mostly} complimentary set of assumptions.  A notable exception here is that Eq.~\ref{eq:floating_system} requires electrons to be present in the sheath whereas Eq.~\ref{eq:alpha_scaling} assumes that the sheath is electron free.

Equations~\ref{eq:alpha_scaling} and~\ref{eq:floating_system} can then solved simultaneously, e.g., using a least-squares solver, which provides both $T_{eV}$ and $V_{plasma}$ without biasing.  

We note that this approach relies on the experimentally determined scaling factor $\alpha \approx 0.74$ discussed in Sec.~\ref{subsec:biased_pip_sheath_thick}, for which a complete physical explanation remains an open question. 
Additionally, while the transitional sheath model is adopted here for the ion current in Eq.~\ref{eq:current_balance}, alternative formulations~\cite{Lobbia2017} such as thin-sheath or OML limits may be substituted when more appropriate.

Table~1 provides a side-by-side comparison of LP and floating PIP measurements of plasma density, sheath thickness, electron temperature,  plasma potential and $r_{probe}/\lambda_D$ across the five discharge currents studied. 
For the LP, the Child--Langmuir sheath thickness $t_{CL}$ was calculated using the density and electron temperature obtained from the LP analysis, and both the unscaled and scaled predictions ($\alpha t_{CL}$) are shown alongside the PIP-measured sheath thickness $t_\mathrm{pip}$. 

Consistent with the results of Sec.~\ref{subsec:biased_pip_sheath_thick}, the PIP sheath thickness agrees more closely with the scaled prediction $\alpha t_{CL}$ than with the unscaled Child--Langmuir model. 
Electron temperature and plasma potential estimates from the floating PIP are also in reasonable agreement with LP measurements, though larger discrepancies emerge at higher discharge currents, which is a similar trend as the density measurements in Sec~4.3.
Overall, these results show that the floating PIP reproduces the principal trends observed with the LP and CL model while additionally providing access to electron damping, a quantity not directly available from conventional LP analysis. 
Taken together, the comparison supports the floating PIP as a viable alternative to the LP for extracting core plasma parameters, with the added advantage of avoiding DC biasing and therefore reducing plasma perturbations.

\begin{table*}[t]
	\centering
	\caption{Comparison of plasma parameters measured using a Langmuir probe (LP) and the floating plasma impedance probe (PIP-float) across five discharge currents.}
	\label{tab:lp_pip_float_compare}
	
	\renewcommand{\arraystretch}{1.25}
	\setlength{\tabcolsep}{5pt}
	
	\begin{tabular}{c|rr|rrr|rr|rr|rr}
		\toprule
		& \multicolumn{2}{c|}{\begin{tabular}[c]{@{}c@{}}Density\\[-2pt]$(10^{14}\,\mathrm{m}^{-3})$\end{tabular}}
		& \multicolumn{3}{c|}{\begin{tabular}[c]{@{}c@{}}Sheath thickness \\at $V_\mathrm{float}$ (mm)\end{tabular}}
		& \multicolumn{2}{c|}{\begin{tabular}[c]{@{}c@{}}$T_{eV}$\\[-2pt](eV)\end{tabular}}
		& \multicolumn{2}{c|}{\begin{tabular}[c]{@{}c@{}}$V_{plasma}$\\[-2pt](V)\end{tabular}}
		& \multicolumn{2}{c}{\begin{tabular}[c]{@{}c@{}}$r_\mathrm{probe}/\lambda_D$\\[-2pt](-)\end{tabular}} \\
		\midrule
		\hline
		$I_D$ (A)
		& \multicolumn{1}{c}{LP} & \multicolumn{1}{c|}{PIP}
		& \multicolumn{1}{c}{$t_{CL}$} & \multicolumn{1}{c}{$\alpha t_{CL}$} & \multicolumn{1}{c|}{$t_\mathrm{pip}$}
		& \multicolumn{1}{c}{LP} & \multicolumn{1}{c|}{PIP}
		& \multicolumn{1}{c}{LP} & \multicolumn{1}{c|}{PIP}
		& \multicolumn{1}{c}{LP} & \multicolumn{1}{c}{PIP} \\
		\midrule
		6  & 1.21 & 1.09 & 5.00 & 3.60 & 3.02 & 2.20 & 2.52 & 23.2 & 23.5 & 12.2 & 11.2 \\
		8  & 1.87 & 1.94 & 3.81 & 2.72 & 2.57 & 2.30 & 3.21 & 24.3 & 25.1 & 15.4 & 13.3 \\
		10 & 2.20 & 2.50 & 3.34 & 2.39 & 2.16 & 2.27 & 2.88 & 22.1 & 22.5 & 16.8 & 15.9 \\
		15 & 4.00 & 5.76 & 2.49 & 1.78 & 1.61 & 3.00 & 3.61 & 19.9 & 22.8 & 19.8 & 21.6 \\
		20 & 4.85 & 7.63 & 2.18 & 1.56 & 1.52 & 2.97 & 4.26 & 18.5 & 25.6 & 21.8 & 22.9 \\
		\bottomrule
	\end{tabular}
	
	\vspace{4pt}
	\footnotesize
	\textit{Notes:} $t_{CL}$ is calculated from LP-derived $n_e$ and $T_{eV}$; $\alpha$ is the empirically determined scaling factor (Sec.~\ref{subsec:biased_pip_sheath_thick}). $t_\mathrm{pip}$ is extracted from the floating-PIP model fit.
\end{table*}

\section{Conclusions  \label{sec:conclusions}}
	
	The first major result of this work is the demonstration that biasing the plasma impedance probe (PIP), a diagnostic traditionally operated in an electrically floating configuration, enables sheath thickness measurements that are in close agreement with Child–Langmuir scaling. Across the range of discharge currents studied, the biased PIP measurements are reconciled with the Child–Langmuir model using a single, consistent correction factor ($\alpha \approx 0.74$). While the physical origin of this factor is not yet fully understood, plausible explanations include differences in sheath geometry, treatment of the presheath, and systematic differences in density determination. Comparisons between PIP- and Langmuir-probe-derived densities show good agreement at lower discharge currents, with larger discrepancies emerging at higher currents. Possible contributors to these discrepancies include known limitations of Langmuir probe (LP) analysis under non-ideal conditions (e.g., non-Maxwellian electron populations or LP's sensitivity to outgassing) and spatial gradients.
	
	The second major result builds on this finding by leveraging the empirically determined $\alpha$ factor and its connection to the Child–Langmuir sheath model to extend the floating PIP analysis. In addition to established measurements of plasma density, sheath thickness, and electron damping, the floating PIP is shown to provide estimates of electron temperature and plasma potential without the need for DC biasing. These results exhibit reasonable agreement with LP measurements across the discharge conditions examined. The key advantage of this development is that temperature and plasma potential can now be accessed using an electrically floating probe, which is inherently less perturbative to the plasma than biased diagnostics.
	
	Finally, this work highlights the complementary nature of LPs and PIPs. LPs directly interrogate the electron energy distribution function through current–voltage characteristics and are therefore well suited for measuring electron temperature and identifying non-Maxwellian features. In contrast, the PIP measures the collective electromagnetic response of the plasma, making it particularly robust for determining plasma density, sheath thickness, and electron damping while requiring fewer assumptions about the detailed shape of the electron distribution. 
	When combined, however, a hybrid LP–PIP diagnostic enables cross-validation and a more complete characterization of the plasma than either technique can provide independently. The results presented here illustrate how combining RF and DC probe concepts can reduce overall model dependence while leveraging the natural strengths of each diagnostic.

\begin{acknowledgments}
	We wish to thank the U.S. Naval Research Laboratory for funding this work. Richeek Dutta acknowledges that his student research was supported by the Naval Research Enterprise Internship Program (NREIP) under the Office of Naval Research (ONR) contract N00014-21-D-4002.  We also wish to thank Dr. Marcel Georgin for his helpful discussions on the PIP and its interaction with the sheath.
\end{acknowledgments}

\section*{Data Availability Statement}
The data (LP and calibrated PIP) and associated code used to generate Figs.~\ref{fig:biased_pip_spectra},~\ref{fig:tpip_vs_tcl}, and~\ref{fig:density_damping} and Table~\ref{tab:lp_pip_float_compare} are hosted on Zenodo~\cite{Brooks2026}.

\bibliography{bibfile}

\end{document}